\documentclass[fleqn,usenatbib]{mnras}

\usepackage{newtxtext,newtxmath}

\usepackage[T1]{fontenc}

\usepackage{threeparttable} 

\DeclareRobustCommand{\VAN}[3]{#2}
\let\VANthebibliography\thebibliography
\def\thebibliography{\DeclareRobustCommand{\VAN}[3]{##3}\VANthebibliography}

\usepackage{graphicx}	
\usepackage{amsmath}	

\usepackage{comment}
\usepackage{color}				
\usepackage[dvipsnames]{xcolor} 
\usepackage{graphicx}	
\usepackage[commandnameprefix=always,markup=underlined]{changes}
\usepackage{subcaption}
\definecolor{wildwatermelon}{rgb}{0.99, 0.42, 0.52}
\definecolor{trolleygrey}{rgb}{0.5, 0.5, 0.5}
\definecolor{cobalt}{rgb}{0.0, 0.28, 0.67}
\definecolor{bleudefrance}{rgb}{0.19, 0.55, 0.91}
\definecolor{cerulean}{rgb}{0.0, 0.48, 0.65}
\definecolor{mspgreen}{rgb}{0.13, 0.55, 0.13}
\definecolor{orange}{rgb}{0.6,0.4,0.0}
\definecolor{denim}{rgb}{0.73, 0.2, 0.52}

\usepackage{soul} 

\newcommand{\igr}{\mbox{IGR~J17014-4306}}
\definecolor{denim}{rgb}{0.73, 0.2, 0.52}

\title[A micronova burst in the IP \igr]{A micronova burst in the intermediate polar \igr}

\author[A. S. Oliveira et al.]{
Alexandre S. Oliveira,$^{1}$\thanks{E-mail: alexandre@univap.br (ASO)}
D. C. Souza,$^{1}$
G. J. M. Luna$^{2,3}$
and C. V. Rodrigues$^{4}$
\\
$^{1}$ IP\&D, Universidade do Vale do Paraíba, 12244-000, São José dos Campos, SP, Brazil\\
$^{2}$ Universidad Nacional de Hurlingham (UNAHUR). Laboratorio de Investigación y Desarrollo Experimental en Computación, Grupo de Astrofísica, \\
Av. Gdor. Vergara 2222, Villa Tesei, Buenos Aires, Argentina \\
$^{3}$ Consejo Nacional de Investigaciones Científicas y Técnicas (CONICET), Argentina.\\
$^{4}$ Instituto Nacional de Pesquisas Espaciais (INPE/MCTI), Av. dos Astronautas, 1758, São José dos Campos, SP, Brazil
}

\date{Accepted XXX. Received YYY; in original form ZZZ}

\pubyear{\the\year{}}

\begin{document}
\label{firstpage}
\pagerange{\pageref{firstpage}--\pageref{lastpage}}
\maketitle

\begin{abstract}

We report the detection of a short optical burst in TESS data of \igr, the eclipsing intermediate polar with the longest known orbital period.
The burst lasts 1.56\,d and shows multiple peaks, reaching $(9.3 \pm 0.2) \times 10^{33}$ erg s$^{-1}$, and releases a total energy of $(3.25 \pm 0.01) \times 10^{38}$ erg. The burst parameters are consistent with those of a micronova eruption, currently understood as a thermonuclear runaway in the magnetically-confined accretion column. From its energy, we infer a burned column mass of $\sim 1.6 \times 10^{-11}$ M$_\odot$, which implies a recurrence time of $\sim 20$ d.
Our search for similar events in long-term Gaia, ASAS-SN, and AAVSO light curves reveals 16 possible fast brightenings over $\sim 11$ yr, suggesting that micronova events may be frequent in \igr. 
Timing analysis of the TESS data shows that the white-dwarf spin period remains stable before and after the burst. During the burst, however, the power spectrum becomes more complex and exhibits multiple peaks.
The classification of \igr\ as a micronova brings the total number of confirmed systems to eight. Its extreme orbital period and eclipsing nature make it an ideal test-bed for further studies of magnetically confined thermonuclear burning on white dwarfs.

\end{abstract}




\section{Introduction}

Cataclysmic Variables (CVs) are binary stellar systems in which a white dwarf accretes material from a low-mass, main-sequence or slightly evolved companion through Roche lobe overflow \citep{1995CAS....28.....W}. Most often, this process takes place through an accretion disc that forms around the white dwarf. In systems where the white dwarf possesses a sufficiently strong magnetic field, the ionized accretion stream couples to the magnetic field lines. This interaction channels material directly onto the white-dwarf's surface near its magnetic poles via accretion columns or curtains. These magnetic Cataclysmic Variables (mCVs) are subdivided into two classes. Polars, characterized by high magnetic-field strengths 
(B$\sim7-230$ MG), exhibit synchronized white-dwarf spin and orbital periods due to magnetic locking. In contrast, Intermediate Polars (IPs), with weaker fields (B$\sim1-10$ MG), are not synchronized; their distinct spin, beat, and orbital periods may be detectable in optical and X-ray emissions. IPs usually possess an accretion disk that is truncated at its inner radius by the white-dwarf's magnetosphere.

Accretion disks in CVs can undergo recurrent thermal instabilities known as Dwarf Nova (DN) outbursts, explained in terms of the disk instability model \citep[DIM,][]{2001NewAR..45..449L}. DN outbursts increase the system's luminosity by 3 -- 6 magnitudes in timescales as short as few days, with recurrence intervals of weeks or months. In IPs, however, these events are scarce and seem to be shorter than in (full-disc) non-magnetic CVs \citep{2017A&A...602A.102H}. 
Furthermore, the short outbursts observed in IPs cannot be understood in the DIM scenario.

However, DN outbursts are not the only type of rapid bursts registered in mCVs.
Donor star flares and the recently identified micronovae and magnetic-gating bursts share similar observational characteristics, making their proper classification difficult. 
Micronovae have been proposed by \citet{2022Natur.604..447S} to explain the rapid succession of bursts observed in the IPs TV Col and EI UMa. They are best-explained as thermonuclear runaways on a small portion of the white-dwarf surface, similar to the type-I X-ray bursts common in X-ray binary neutron stars.
In contrast to classical nova eruptions, which are thermonuclear runaways spread over the entire white-dwarf surface, micronovae are restricted to the base of the accretion column by magnetic confinement of the accretion stream \citep{2022MNRAS.514L..11S}. 

In the magnetic-gating model \citep{2012MNRAS.420..416D,2017Natur.552..210S}, accretion from the donor star is trapped at the disc's co-rotation radius by a centrifugal barrier created by the rotating magnetosphere of the white dwarf. When the increasing pressure eventually allows the accumulated material to cross the barrier, the material is discharged onto the white dwarf, and a short burst of accretion energy is released. Currently, the identified members of the magnetic-gating group are MV Lyr \citep{2017Natur.552..210S}, TW Pic \citep{2022NatAs...6...98S}, V1223 Sgr \citep{2022A&A...664A...7H}, V1025 Cen \citep{2022ApJ...924L...8L} and CTCV J0333-4451 \citep{2024ApJ...962L..34I}.

Alternatively, donor stellar flares were detected in very few mCVs, all of them of the polar type: AM Her \citep{1993AnIPS..10..237S}, V358 Aqr \citep{2017A&A...603A..47B}, and MQ Dra \citep{2021MNRAS.504.4072R}, with TESS (Transiting Exoplanet Survey Satellite) observations available only for the latter.

\igr\ was identified in hard X-rays as a possible mCV by \citet{2013A&A...556A.120M}. Observations from XMM-Newton \citep{2017MNRAS.470.4815B} revealed total X-ray eclipses and partial eclipses in the optical B band, alongside modulation\sout{s} of 1859~s corresponding to the white-dwarf spin. 
\igr\ has an orbital period of 12.8 hours, making it the eclipsing IP with the longest known orbital period.

\cite{2017Natur.548..558S} proposed that \igr\ is the counterpart of the historical Nova Sco 1437, a fast-declining classical nova. The same study also identified three transient brightenings in the DASCH\footnote{DASCH - Digital Access to a Sky Century @ Harvard} optical light curve, occurring in 1934, 1935, and 1942. During these events, the system brightened from a quiescent magnitude of $\sim$17 to approximately 12 in 1934 and to 14.5--15 in the subsequent eruptions. These brightenings were interpreted as dwarf-nova outbursts. The proposed connection to Nova Sco 1437 has since been challenged by \citet{2025A&A...694A.105G}, who concluded that the ionized mass of the nebula and its expansion velocity are inconsistent with those expected for a nova remnant. Instead, they argue that the nebula is likely an evolved planetary nebula, unrelated to the remnant of Nova Sco 1437, and that \igr\ is by chance projected onto the nebula.

The IP nature of \igr\ was corroborated by \citet{2018MNRAS.473.4692P}, who detected spin-modulated circular polarization. The stability of the polarization sign, despite variations in intensity, implies that only one magnetic pole of the white dwarf is observed. A decrease in circular polarization during  the eclipse is likely caused by the occultation of the cyclotron emission region. More recently, by analysing TESS and AAVSO eclipse timings, \cite{2025ApJ...988...87Z} claimed that the orbital period of \igr\ is increasing at a remarkably high rate, comparable to that observed in compact binary supersoft X-ray sources (CBSS) and recurrent novae.

In this paper, we present the detection and analysis of a rapid burst observed in \igr\ by TESS in June 2025.
Section~\ref{sec-2} describes the TESS data and calibration. Section~\ref{sec-3} presents the results of the burst analysis, and Section~\ref{sec-4} provides the discussion and conclusions.

\section{Observations}\label{sec-2}

TESS \citep{2015JATIS...1a4003R} observed \igr\ in sectors 12, 39, 66, and 93 (in 2019, 2021, 2023, and 2025, respectively), all with a cadence of 120 s. TESS data were processed by the Science Processing Operations Center (SPOC) pipeline and were downloaded from the Mikulski Archive for Space Telescopes\footnote{\url{https://mast.stsci.edu/portal/Mashup/Clients/Mast/Portal.html}} (MAST) using the Python package \textsc{Lightkurve} \citep{Lightkurve2018}. We used the Simple Aperture Photometry (SAP) flux and discarded all data points with a quality flag greater than 0. Timestamps are expressed in Barycentric Tess Julian Date (BTJD), deﬁned as BTJD = BJD - 2457000.0, where BJD stands for Barycentric Julian Date in Barycentric Dynamical Time.

ASAS-SN \citep{2017PASP..129j4502K} observed \igr\ from March 2016 to September 2018 in filter V, and from June 2018 to October 2025 in filter g, with a cadence of three data points per visit. Data were downloaded from the Sky Patrol server\footnote{\url{https://asas-sn.osu.edu/}}, in the Aperture Photometry mode. ASAS-SN data timestamps are given in Heliocentric Julian Date (HJD).

Gaia \citep{2016A&A...595A...1G} observed \igr\ from September 2014 to March 2017 in $G$, $G_{BP}$ and $G_{RP}$ bands. The DR3 data were downloaded from the Gaia Archive\footnote{\url{https://gea.esac.esa.int/archive/}}. Gaia observation times are expressed in units of BJD in TCB (Barycentric Coordinate Time) minus 2455197.5, corresponding to a reference time $T_0$ at 2010-01-01T00:00:00.

AAVSO International Database has \igr\ data from September 2016 to July 2025. Data in V and CV (CV stands for unfiltered observation calibrated with V-band comparison star) bands were downloaded from the Light Curve server\footnote{\url{https://www.aavso.org/LCGv2/}}.

\subsection{TESS data calibration}

We converted the TESS fluxes (e$^-$s$^{-1}$) in sector 93  to luminosity (erg s$^{-1}$) by calibrating the TESS flux with ASAS-SN photometry and using the distance to the source.
The procedure we adopted was the same one presented in \citet{2022Natur.604..447S} to achieve homogeneous results with the literature. We started by correlating TESS data with simultaneous ASAS-SN g-band photometry. The data points to be correlated consisted of those taken within 120 s of each other. No colour-term variations were considered. This was done separately in each half of the TESS sector, limited by the gap created by TESS data downlink. The correlation yields the following linear transformation between TESS e$^-$s$^{-1}$ flux and ASAS-SN mJy spectral flux density \citep{2024MNRAS.529..664V}: 

\begin{equation}
F_{ASAS-SN}~[mJy] = A~ \times ~F_{TESS}~ [e^- s^{-1}] + C,    
\end{equation}

\noindent where A and C are free parameters. For the first half of sector 93 the parameters are $A = 0.0156 \pm 0.0019 ~ \mathrm{mJy}/e^- \mathrm{s}^{-1}$ and $C=-2.6022 \pm 0.5064$~mJy, while for the second half $A = 0.0148 \pm 0.0021 ~ \mathrm{mJy}/e^- \mathrm{s}^{-1}$ and $C=-2.4902 \pm 0.5519$~mJy. 

\begin{figure*}
	\includegraphics[scale=0.6]{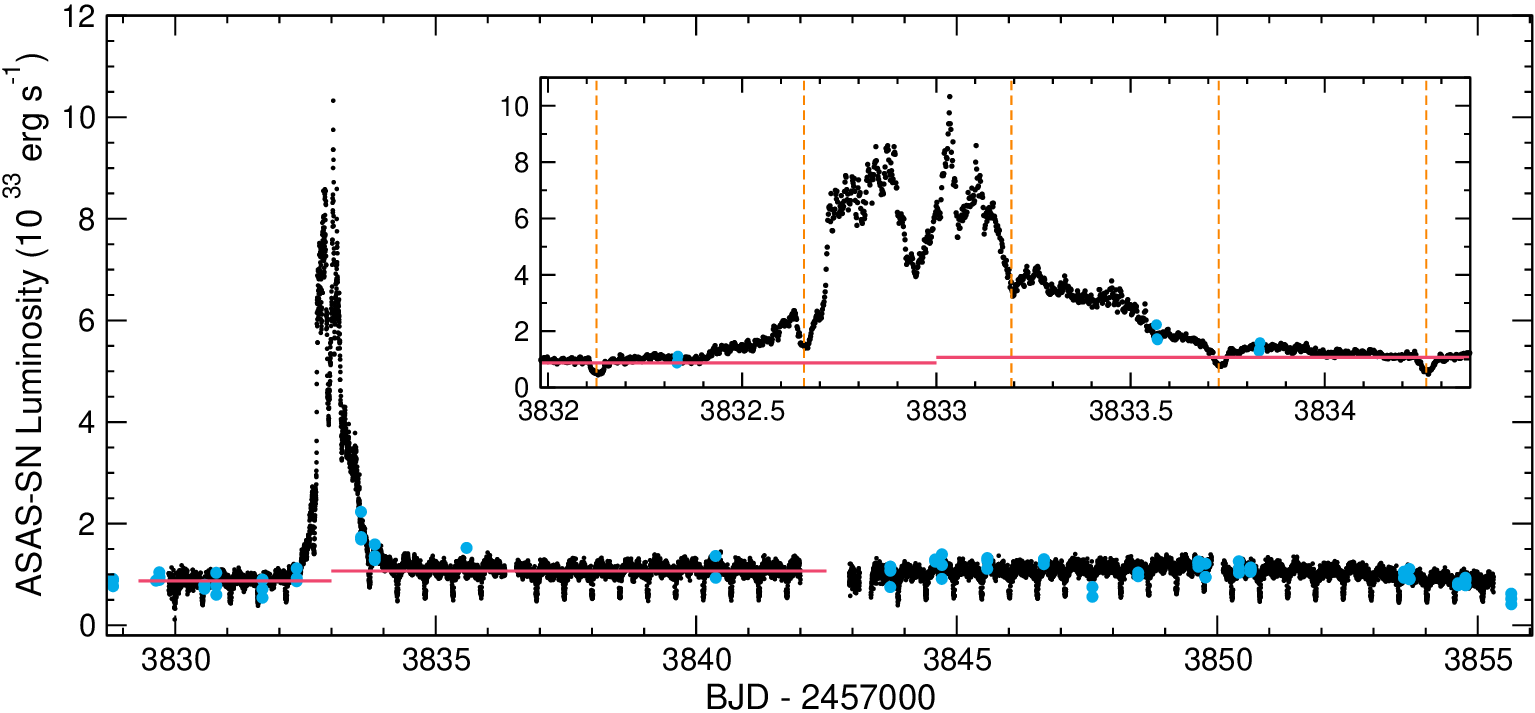}
    \caption{TESS sector 93 light curve of \igr\ (black dots), with the flux scale converted to luminosity using the Gaia distance to the source. The blue points represent the ASAS-SN g-band data used in the calibration, and the pink horizontal lines show the mean quiescent luminosity before and after the burst. The inset figure depicts an expanded view around the burst. The vertical orange dashed lines represent the eclipse timings calculated from the linear ephemeris of \citet{2017Natur.548..558S}.}
    \label{figTESS}
\end{figure*}

Assuming the Gaia distance to \igr\ as $988_{-36}^{+37}$~pc \citep{2021AJ....161..147B}, we converted the spectral flux densities into luminosities. Fig.\ref{figTESS} shows the TESS sector 93 light curve converted to luminosity and the ASAS-SN g-band points used in the correlation. This luminosity is a lower limit, since no bolometric correction have been applied in the cross-calibration.

\section{Results}\label{sec-3}

\subsection{The burst energy}

The burst detected in the TESS sector 93 light curve presents multiple peaks (Fig.~\ref{figTESS}), similar to those observed in TV Col and EI UMa \citep{2022Natur.604..447S}, CP Pup \citep{2024MNRAS.529..664V}, and PBC~J0801.2–4625 \citep{2024MNRAS.530.3974I}. The two main peaks are separated by 0.26 d, while the burst duration is 1.56 d. The eclipses are clearly visible, even during the burst. No precursors are observed, in contrast to what is reported for the bursts in TV Col and \mbox{ASASSN-19bh} \citep{2022Natur.604..447S}. 
The data indicate that after the burst, the system dropped to a luminosity level ($1.07~\times~10^{33}$~erg~s$^{-1}$) that is slightly higher than before the burst ($0.87~\times~10^{33}$~erg~s$^{-1}$). Non-parametric statistical tests (Mann-Whitney U test and permutation test for medians) indicate that the difference is statistically significant. Note that this comparison between the luminosity levels considers only the first half of the sector (i.e. before the gap), so it is not subject to count rate offsets due to TESS data downlink.

To determine the energy of the burst, we integrated the calibrated luminosity curve, after subtracting a quiescent emission level of $0.97~\times~10^{33}$~erg~s$^{-1}$ estimated as the mean of the pre- and post-burst quiescent values. The total (integrated) energy obtained is $(3.25\pm0.01)~\times~10^{38}$~erg, while the peak luminosity is $(9.3\pm0.2)~\times~10^{33}$~erg~s$^{-1}$. As mentioned above, these should be considered as lower limits.

\subsection{The burst classification}

\citet{2024ApJ...962L..34I} proposed empirical diagnostic diagrams based on the integrated energies, peak luminosities, durations, and recurrence frequencies to differentiate the observational signatures of the various types of rapid optical bursts observed in CVs. Their results demonstrate that dwarf-nova eruptions, micronovae, magnetically gating bursts, and donor flares occupy separate regions in the parameter space, indicating that these phenomena originate from different physical processes. 

\begin{figure*}
		\includegraphics[width=\textwidth]{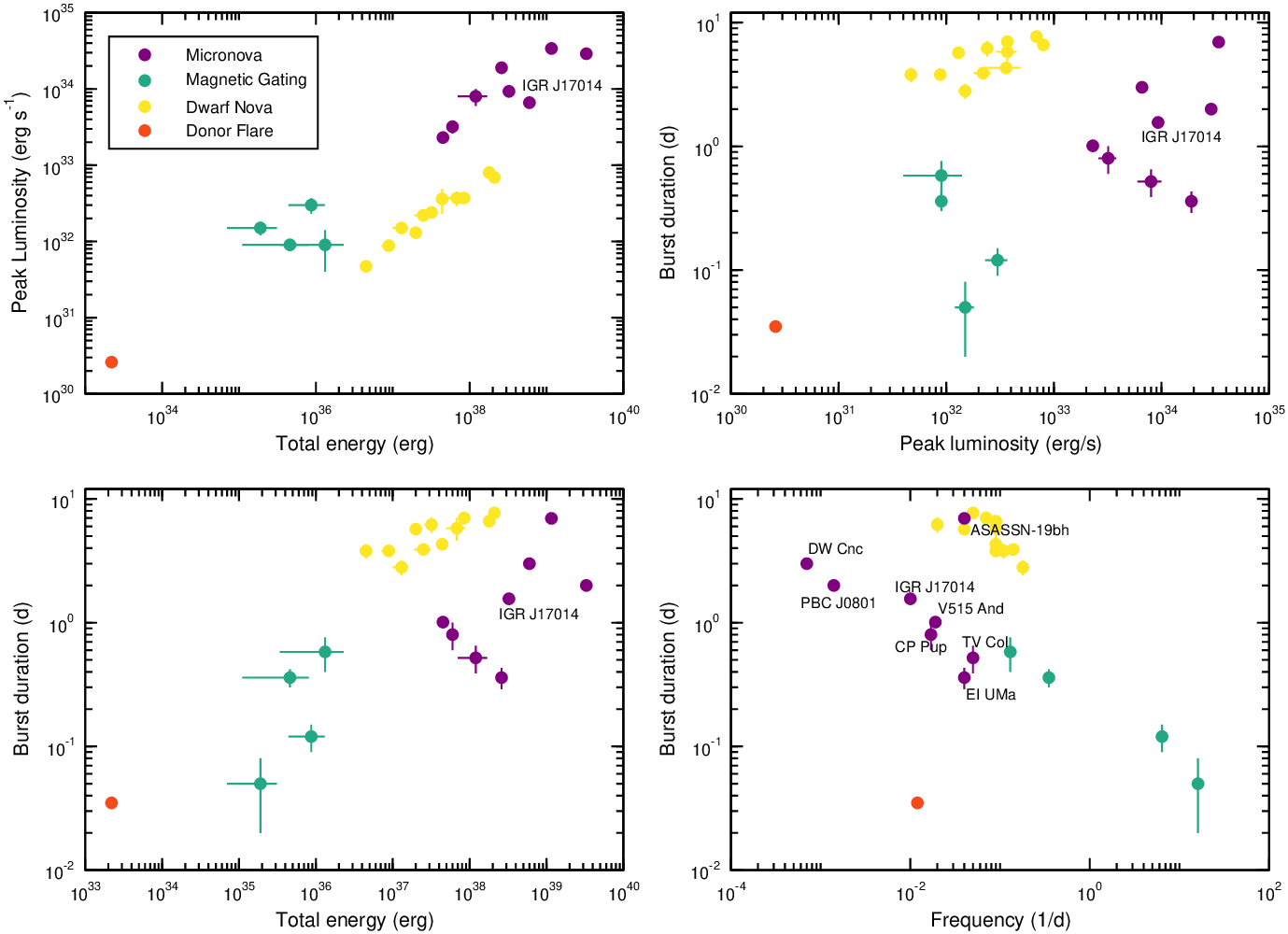}
    \caption{Burst properties of CVs as adapted from Figure 2 of \citet{2024ApJ...962L..34I}. }
    \label{figdiagnostic}
\end{figure*}

Following the prescription from \citet{2020ApJ...890...46T} and \citet{2024ApJ...962L..34I}, the burst frequency in \igr\ is 0.01 d$^{-1}$, calculated as the ratio of the number of bursts to the continuous monitoring period.
Table~\ref{tab:bursts_properties} lists the bursts peak optical luminosity, total optical energy, duration, and frequency for \igr\ and the known sample of micronovae, while Fig.~\ref{figdiagnostic} presents these objects in the diagnostic diagrams alongside the magnetic-gating, donor flare, and dwarf novae systems from \citet{2024ApJ...962L..34I}. The diagrams show that the burst detected in \igr\ in June 2025 is consistent with a micronova eruption.

\begin{table*}
\begin{threeparttable}[b]
\centering
\caption{Properties of the bursts observed in the known sample of micronovae.}
\label{tab:bursts_properties}
\begin{tabular}{lccccc} 
\hline
Object   & Peak Luminosity  & Total  Energy  &  Duration  & Frequency  & Reference\\
         & ($\times 10^{34}$ erg s$^{-1}$) & ($\times 10^{38}$ erg) & (d) & (d$^{-1}$) & \\
\hline
TV Col & $0.8\pm0.2$ & $1.2\pm0.5$ & $0.52\pm0.13$ & 0.05 & (1), (2) \\
EI UMa & $1.9\pm0.2$ & $2.6\pm0.3$ & $0.36\pm0.07$ & 0.04 & (1), (2)  \\
ASASSN-19bh & $3.4$ & $11.6$ & 6.96 & 0.04 & (1), (2) \\ 
CP Pup & $0.32\pm0.06$ & $0.6$ & $0.8\pm0.2$ & 0.017 & (1), (3) \\
PBC J0801 & $2.9$ & $33$ & 2 & 0.0014 & (4) \\ 
DW Cnc & $0.66$ & $6$ & 3 & 0.0007  & (5) \\ 
V515 And & $0.23\pm0.03$ & $0.45\pm0.01$ & 1.010 & 0.019  & (6) \\
\igr &  $0.93\pm0.02$ & $3.25\pm0.01$ & 1.56 & 0.01 & (7) \\   
\hline
\end{tabular}
\begin{tablenotes}
   \item []    \textbf{References:} (1) \citet{2024ApJ...962L..34I}; (2) \citet{2022Natur.604..447S}; (3) \citet{2024MNRAS.529..664V}; (4) \citet{2024MNRAS.530.3974I}; (5) \citet{2025MNRAS.539.2424V}; (6) \citet{2026MNRAS.tmp..321R}; (7) this work.
\end{tablenotes}
\end{threeparttable}
\end{table*}

\subsection{Previous bursts}\label{3.3}

\begin{figure*}
    \centering
    
    \begin{subfigure}{1.0\textwidth}
        \centering
        \includegraphics[width=0.8\linewidth]{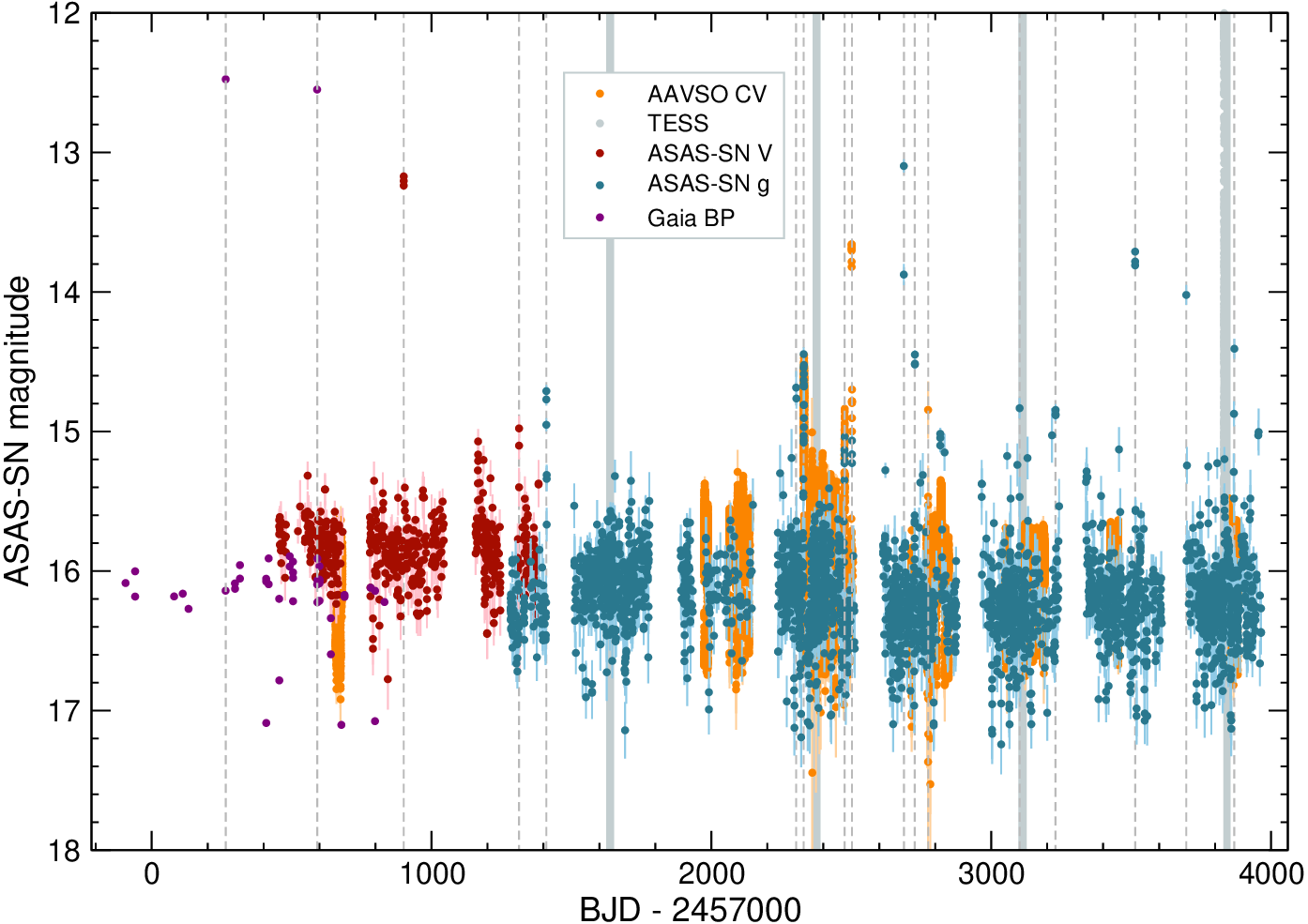}
    \end{subfigure}
    
    
    \begin{minipage}{1.0\textwidth}
        \centering
        \begin{subfigure}{0.4\textwidth}
            \centering
            \includegraphics[width=\linewidth]{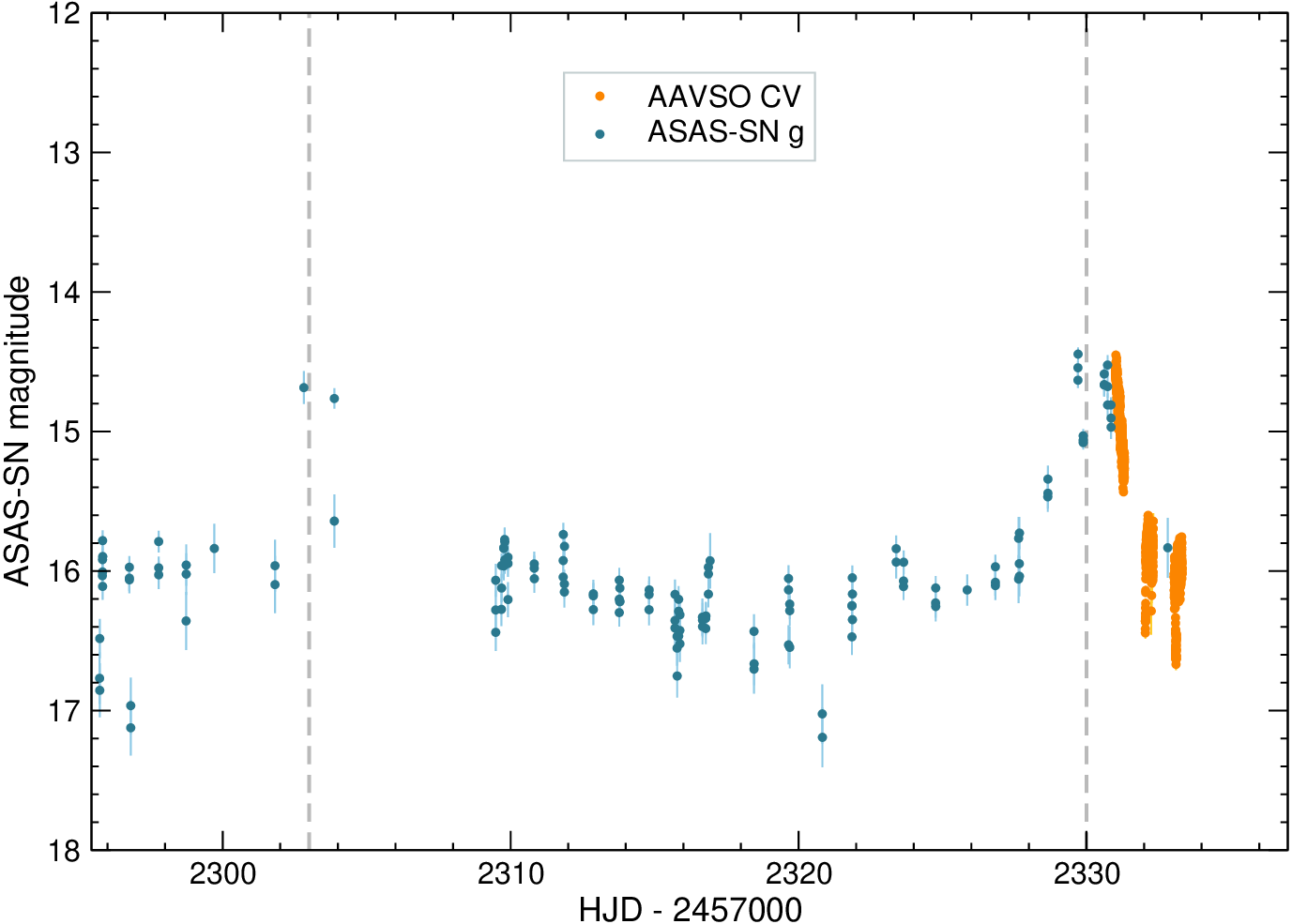}
        \end{subfigure}
        \begin{subfigure}{0.4\textwidth}
            \centering
            \includegraphics[width=\linewidth]{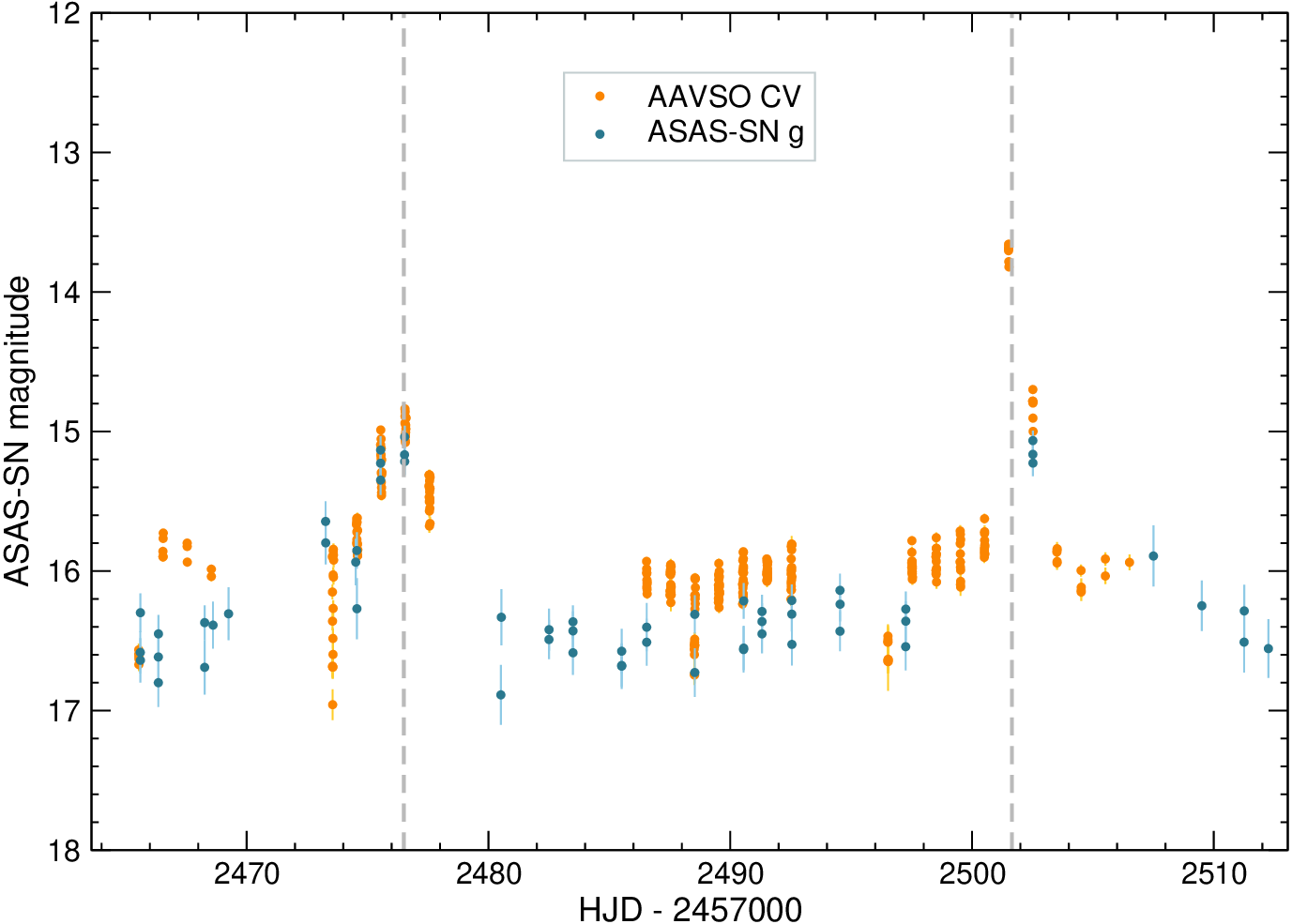}
        \end{subfigure}
    \end{minipage}
    
    \caption{Long-term optical light curve of \igr\ combining data from AAVSO (CV band, orange), ASAS-SN (V band, dark red; $g$ band, light blue), Gaia ($G_\mathrm{BP}$ band, purple), and TESS (gray shaded regions marking the four observed sectors). 
    Vertical dashed lines mark sixteen possible micronova events. The lower panels show zoomed-in views of bursts simultaneously covered by ASAS-SN $g$-band and AAVSO data, illustrating their short durations (less than 4\,d). }
    \label{figLCs}
\end{figure*}

Except for sector 93, none of the preceding TESS sectors (12, 39, or 66) displays any hint of bursts in \igr. 
To evaluate the occurrence of other bursts, we created a long-term light curve using Gaia, ASAS-SN, and AAVSO data (Fig.~\ref{figLCs}).
The average brightness in quiescence fluctuates between $\sim15.5$ and $\sim16.5$ magnitudes.
There are many points consistent with bursts, many of which are not isolated, and the brightest one reaches around 12.5~mag. 
We arbitrarily selected the data points brighter than 15 mag, resulting in 16 possible events (marked as dashed vertical lines in Fig.~\ref{figLCs}). The burst at BJD 2457900 was also detected is the ASAS-SN data by \citet{2026MNRAS.549ag822M}, who interpreted it as an evidence of a short outburst but not a definitive micronova identification. Due to the data sampling, the Gaia, ASAS-SN, and AAVSO light curves are prone to miss bursts. This is exactly the case of the burst detected in TESS sector 93 (see Fig.~\ref{figTESS}): ASAS has a relatively good coverage during this sector, but it was not able to detect the burst. The lower panels of Fig.~\ref{figLCs} detail four bursts, some of which are simultaneously covered by ASAS-SN and AAVSO data. These selected bursts last less than 4 d, with recurrence times as short as 27 days, and the burst peaks are possibly missing.

\subsection{The white-dwarf spin period}

\cite{2025ApJ...988...87Z} studied the behaviour of the white-dwarf spin using AAVSO and TESS data from Sectors 12, 39, and 66, observed prior to Sector 93, in which we detected the burst. They found that the spin-pulse profile varies, whereas the spin period shows no significant change. We analysed TESS Sector 93 data to investigate whether the micronova burst caused any change in the spin period. 

We divided the first half of Sector 93 light curve into three segments of equal duration. The first and third segments correspond to the epochs immediately before and after the burst, respectively, while the second segment covers nearly the entire burst. The previous TESS sectors and the second half of Sector 93, following the data downlink gap, were also analysed. All segments were detrended using a Savitzky–Golay filter to remove orbital and longer-term variations, and the residuals were analysed with Fourier techniques over a broad frequency range ($20-70$~d$^{-1}$). The Lomb–Scargle periodogram of all segments, excluding the one containing the burst, show a single prominent peak at the spin frequency of $46.48$~d$^{-1}$ ($\sim1858.9$~s), with power significantly above the false alarm probability (FAP) level. This indicates that the spin period remained stable, within uncertainties, after the micronova eruption. However, in Segment 2 (i.e. during the eruption), the power spectrum becomes more complex and exhibits multiple peaks, all with comparatively lower power than those outside the burst interval. The slightly most prominent peak is at $26.06$~d$^{-1}$, followed by another at $41.44$~d$^{-1}$ and by a peak near, but slightly above, the spin frequency at $46.56$~d$^{-1}$. 

This enrichment of the power spectrum during the eruption and its return to its pre-eruption state afterward was also described for the eruption observed in the intermediate polar PBC~J0801.2--4625 by \citet{2024MNRAS.530.3974I}. Those authors considered micronovae and dwarf novae as possible eruption mechanisms for PBC J0801.2–4625 and concluded that the observed eruption characteristics are more consistent with a micronova. In this context, they suggested that variations in the white-dwarf spin period could be analogous to burst oscillations (BOs) commonly seen in Type I X-ray bursts, hence the term micronova oscillations.


\begin{figure}
	   
	\includegraphics[width=\columnwidth]{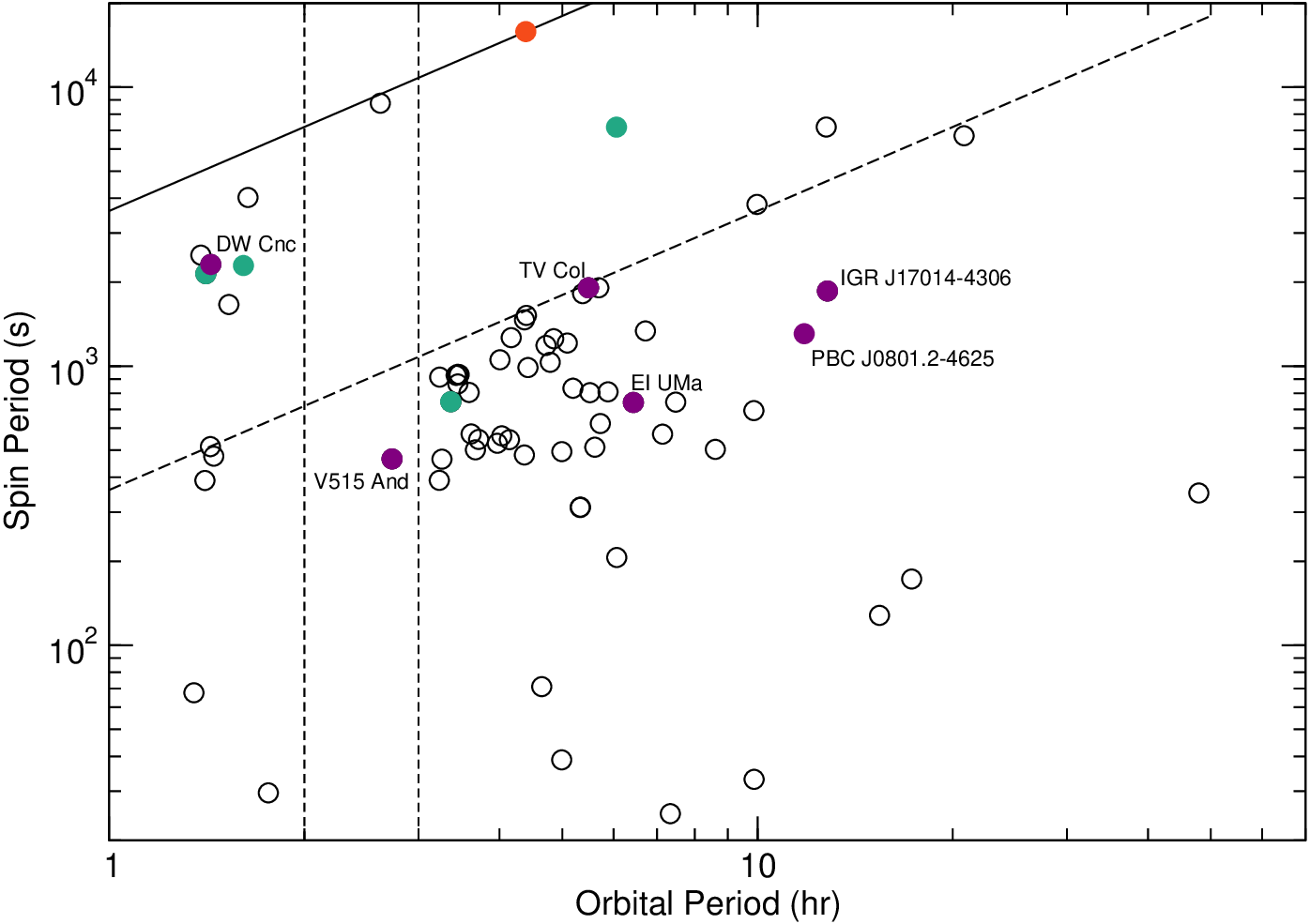}
    \caption{Spin period versus orbital period for mCVs. Micronovae are depicted in purple, magnetic-gating systems in green and the donor flare polar in orange. Empty circles represent confirmed IPs and IP candidates listed in Mukai's homepage (see text). The vertical lines are the period gap interval of 2 to 3~h and the diagonal lines represent $P_{\rm spin}=P_{\rm orb}$ (solid) and $P_{\rm spin}/P_{\rm orb}=0.1$ (dashed).
    }
    \label{fig:pspinporb}
\end{figure}


\section{Discussion and Conclusions}\label{sec-4}

The burst detected in \igr\ during TESS Sector 93 reaches a peak optical luminosity of $(9.3 \pm 0.2) \times 10^{33}\,\mathrm{erg\,s^{-1}}$, a total radiated energy of $(3.25 \pm 0.01) \times 10^{38}\,\mathrm{erg}$, and lasts $1.56$\,d. When placed on the diagnostic diagrams of \citet[see Fig.~\ref{figdiagnostic}]{2024ApJ...962L..34I}, it falls within the region occupied by known micronovae and clearly away from dwarf-nova outbursts, magnetically gated bursts, and donor flares. This provides strong evidence that the event in \igr\ is powered by a localized thermonuclear runaway.

Besides the bursts we detected in the long-term Gaia, ASAS-SN, and AAVSO optical monitoring, \citet{2017Natur.548..558S} identified three transient brightenings (in 1934, 1935, and 1942) in historical DASCH observations, which they interpreted as dwarf-nova outbursts. Given the similarities of the amplitudes, these historical eruptions are plausible micronova events.

Assuming hydrogen burning through the CNO cycle with an energy release of $\sim 10^{16}$~erg~g$^{-1}$ \citep[e.g.][]{1971MNRAS.152..307S, 1972ApJ...176..169S}, the observed burst energy of \igr\ implies a burned column mass of $M_{\rm col} \simeq 1.6\times10^{-11}\,{\rm M_\odot}$. 

We followed the magnetically confined thermonuclear runaway model of \citet{2022MNRAS.514L..11S} to estimate the pressure at the base of the accretion column:

\begin{equation}
    P_{\rm base} = \frac{G M_{\rm WD} M_{\rm col}}{4\pi f R_{\rm WD}^{4}}.
\end{equation}

We adopted a white-dwarf mass $M_{\rm WD} \simeq 0.94^{+0.09}_{-0.07}\,M_\odot$ and a fractional accretion area $f = A_{\rm col}/A_{\rm WD} = 6.3 \times 10^{-4}$ from \cite{Souza2026}. 
The white-dwarf radius, $R_{\rm WD}$, is estimated using the \citet{1972ApJ...175..417N} theoretical mass-radius relationship and the procedure described in \citet{yu2018empirical}, which results in $R_{\rm WD} \simeq 8.51~\rm \times 10^{-3}~R_\odot\,$. 
This yields $P_{\rm base} \simeq 4.3\times 10^{15}\ \mathrm{dyn\,cm^{-2}}$. For a TNR ignition, $P_{\rm base} \simeq P_{\rm crit}$ with $P_{\rm crit} \sim 10^{18}$--$10^{20}\,\mathrm{dyn\,cm^{-2}}$ \citep[][]{2005ApJ...623..398Y,2020A&A...634A...5J,2022Natur.604..447S} requires a critical column mass $M_{\rm col,crit} \simeq 3.75 \times 10^{-9}\,M_\odot$. Thus, our inferred $M_{\rm col}$ is only $\simeq 0.4$ percent of what is needed to reach $P_{\rm crit}$. The discrepancy may reside in the fractional accretion area $f$, as previously pointed out by \citet{2022Natur.604..447S}. As suggested by \citet{2022MNRAS.514L..11S}, other mechanisms such as the Rayleigh--Taylor instability could allow $P_{\mathrm{crit}}$ to be reached with wider $f$, for which further MHD numerical calculations are needed.

Using the inferred burned column mass $M_{\rm col}$ and the secular mass-transfer rate $\dot{M} = 2.9 \times 10^{-10}\,M_\odot\,\mathrm{yr^{-1}}$ from X-ray multi-temperature plasma modelling \citep{Souza2026}, the lower-limit recurrence time is

\begin{equation}
    t_{\mathrm{rec}} \equiv \frac{M_{\mathrm{col}}}{\dot{M}} \approx 0.056 ~ \mathrm{yr} \;\;(\approx 20 ~ \mathrm{days}).
\end{equation}

This estimate is in good agreement with the shortest observed recurrence intervals inferred from the long-term optical monitoring (Fig. \ref{figLCs}), where individual bursts can occur as close as $\sim 26$\,d apart, and with the overall detection of 16 candidate events over about 11 years.

The spin-to-orbital period ratio of $P_{\rm spin}/P_{\rm orb} \approx 0.040$ places this system in the disc-fed accretion regime \citep[see][]{2004ApJ...614..349N, 2008ApJ...672..524N}.
Considering $P_{\rm spin} = 1859$\,s and assuming spin equilibrium between the accretion disc and the white dwarf, we used the empirical spin period–magnetic moment relation for intermediate polars \citep{1989MNRAS.237..715N,1994PASP..106..209P} to obtain a magnetic moment $\mu \simeq 3.59 \times 10^{33}\,\mathrm{G\,cm^{3}}$, which is close to the value found by \citet{2017MNRAS.470.4815B}: $\mu \simeq 5 \times 10^{33}\,\mathrm{G\,cm^{3}}$. 
Adopting the inferred radius for a white dwarf with $0.94\,M_\odot$, the magnetic field is $B \simeq 17^{+7}_{-4}\,\mathrm{MG}$.

To date, almost all confirmed micronovae have been detected in IPs. The only exceptions are ASASSN-19bh and CP Pup, whose precise classifications remain unknown. Figure~\ref{fig:pspinporb} shows the $P_{\rm spin}-P_{\rm orb}$ diagram for confirmed IPs from the catalogue of K. Mukai\footnote{\url{https://asd.gsfc.nasa.gov/Koji.Mukai/iphome/iphome.html}}, adding/highlighting the micronova, magnetic-gating, and donor flare systems. Their distribution in the diagram is random and creates no distinction among the burst classes. The micronova systems are spread across the orbital period range, with objects below, inside, and mostly above the gap -- a distribution that may simply reflect the underlying IP population statistics.   

The micronova trigger mechanism proposed by \citet{2022MNRAS.514L..11S} favours systems with both high white-dwarf mass and high mass-transfer rate, which accelerates the build-up of fuel at the base of the magnetically confined accretion column. For systems above the period gap, angular momentum loss is dominated by magnetic braking, leading to higher secular mass-transfer rates than below the gap, where gravitational radiation dominates. Consistent with this picture, \igr\ is among the IPs with the longest orbital periods, holds the record for the longest orbital period among micronova systems, and harbours a massive white dwarf, $M_{\rm WD}~\simeq~0.94\,M_\odot$ \citep{Souza2026}.
 
The lack of confirmed micronovae in polars is particularly noteworthy. The mass required to trigger a micronova scales with the fractional accretion area $f$, and polars are expected to have the smallest $f$ values among magnetic CVs \citep{2022Natur.604..447S}, which would favour ignition. On the other hand, polars lie predominantly below the period gap (e.g.\ \citealt{2015SSRv..191..111F}), where white-dwarf masses are typically lower and gravitational radiation drives low mass-transfer rates. Such low $\dot{M}$ would lengthen recurrence times, making micronovae in polars more difficult to detect.

The white-dwarf spin behaviour offers further insight into the event's nature. 
During the burst, the power spectrum shows multiple peaks, 
including a signal near but not exactly at the spin frequency. This resembles that reported by 
\citet{2024MNRAS.530.3974I} for PBC~J0801.2$-$4625 and is interpreted as possible micronova 
oscillations, analogous to Type~I X-ray burst oscillations. 
Our analysis of TESS Sector~93 shows that the spin period ($P_{\rm spin} \simeq 1859$~s) remains 
stable before and after the burst. 
This indicates that angular momentum variation during a single micronova is negligible
compared to secular torques on the white dwarf.

In summary, the TESS detection of a short optical burst in \igr, combined with long-term optical monitoring, strongly supports its classification as a recurrent micronova. To date, eight micronovae are confirmed, all of which are probably IPs, and they appear to be homogeneously distributed over the orbital period range. The system thus provides a valuable addition to this still small population and, owing to its long orbital period and eclipsing geometry, offers a promising laboratory for further tests of magnetically confined thermonuclear burning on white dwarfs.

\section*{Acknowledgements}
DCS acknowledges the support of the Coordenação de Aperfeiçoamento de Pessoal de Nível Superior (CAPES), the Universidade do Vale do Paraíba (UNIVAP), and its Research and Development Institute (IP$\&$D). GJML is member of the CIC-CONICET (Argentina).
CVR thanks the Brazilian Space Agency (AEB) for the support from PO 20VB.0009 and the Brazilian National Council for Scientific and Technological Development – CNPq (Proc: 305991/2024-8).
This paper includes data collected by the TESS mission. Funding for the TESS mission is provided
by NASA’s Science Mission Directorate.
This work also makes use of data from the All-Sky Automated Survey for Supernovae (ASAS-SN) project. ASAS-SN is funded by the Gordon and Betty Moore Foundation through grant GBMF5490 and GBMF10501, and by the National Science Foundation through grant AST-2407206. We also thank the Las Cumbres Observatory (LCO) for their ongoing support in hosting the ASAS-SN telescopes.
This work has made use of data from the European Space Agency (ESA) mission
{\it Gaia} (\url{https://www.cosmos.esa.int/gaia}), processed by the {\it Gaia}
Data Processing and Analysis Consortium (DPAC,
\url{https://www.cosmos.esa.int/web/gaia/dpac/consortium}). Funding for the DPAC
has been provided by national institutions, in particular the institutions
participating in the {\it Gaia} Multilateral Agreement. We acknowledge with thanks the variable star observations from the AAVSO International Database contributed by observers worldwide and used in this research.

\section*{Data Availability}

TESS data used in this work are available on the MAST webpage \url{https://mast.stsci.edu/portal/Mashup/Clients/Mast/Portal.html}.
ASAS-SN data \citep{2014ApJ...788...48S,2017PASP..129j4502K} are available on the ASAS-SN webpage \url{https://asas-sn.osu.edu/}, and AAVSO data are available on the AAVSO webpage \url{https://www.aavso.org/LCGv2/}. Gaia data are available on the Gaia Archive at \url{https://gea.esac.esa.int/archive/}.



\bibliographystyle{mnras}
\bibliography{references} 








\bsp	
\label{lastpage}
\end{document}